\documentclass[sigconf,natbib=true]{acmart}

\AtBeginDocument{%
  }

\setcopyright{acmlicensed}
\copyrightyear{2018}
\acmYear{2018}
\acmDOI{XXXXXXX.XXXXXXX}
\acmConference[Conference acronym 'XX]{Make sure to enter the correct
  conference title from your rights confirmation emai}{June 03--05,
  2018}{Woodstock, NY}
\acmISBN{978-1-4503-XXXX-X/18/06}

\usepackage[utf8]{inputenc} 
\usepackage[T1]{fontenc}    
\usepackage{hyperref}       
\usepackage{url}            
\usepackage{amsfonts}       
\usepackage{nicefrac}       
\usepackage{microtype}      
\usepackage{ulem}
\usepackage{appendix}
\usepackage{wrapfig}
\usepackage{xurl}

\usepackage{booktabs}
\usepackage{multirow}
\usepackage{algorithm}
\usepackage{algorithmic}
\usepackage{enumitem}
\usepackage{enumerate}
\usepackage{xcolor}
\usepackage{graphicx}
\usepackage{subcaption}
\usepackage{diagbox} 

\usepackage[most,skins,theorems]{tcolorbox}
\tcbset{
  aibox/.style={
    width=\linewidth,
    top=8pt,
    bottom=4pt,
    colback=blue!6!white,
    colframe=black,
    colbacktitle=black,
    enhanced,
    center,
    attach boxed title to top left={yshift=-0.1in,xshift=0.15in},
    boxed title style={boxrule=0pt,colframe=white,},
  }
}
\newtcolorbox{AIbox}[2][]{aibox,title=#2,#1}
\definecolor{lightblue}{rgb}{0.22,0.45,0.70}

\newcommand{\eg}{\textit{e.g.}}
\newcommand{\ie}{\textit{i.e.}}

\begin{document}

\title{Agentic Information Retrieval}

\author{Weinan Zhang}
\affiliation{%
    \institution{Shanghai Jiao Tong Univeristy}
    \city{Shanghai}
    \country{China}}
\email{wnzhang@sjtu.edu.cn}

\author{Junwei Liao}
\affiliation{%
    \institution{Shanghai Jiao Tong Univeristy}
    \city{Shanghai}
    \country{China}}
\email{jwliao.ai@gmail.com}

\author{Ning Li}
\affiliation{%
    \institution{Shanghai Jiao Tong Univeristy}
    \city{Shanghai}
    \country{China}}
\email{lining01@sjtu.edu.cn}

\author{Kounianhua Du}
\affiliation{%
    \institution{Shanghai Jiao Tong Univeristy}
    \city{Shanghai}
    \country{China}}
\email{kounianhuadu@sjtu.edu.cn}

\author{Jianghao Lin}
\affiliation{%
    \institution{Shanghai Jiao Tong Univeristy}
    \city{Shanghai}
    \country{China}}
\email{chiangel@sjtu.edu.cn}
\renewcommand{\shortauthors}{Weinan Zhang et al.}
\renewcommand{\shorttitle}{Agentic Information Retrieval}

\begin{abstract}
Since the 1970s, information retrieval (IR) has long been defined as the process of acquiring relevant information items from a pre-defined corpus to satisfy user information needs. 
Traditional IR systems, while effective in domains like web search, are constrained by their reliance on static, pre-defined information items. 
To this end, this paper introduces \textbf{\textit{agentic information retrieval}} (\textbf{Agentic IR}), a transformative next-generation paradigm for IR driven by large language models (LLMs) and AI agents. 
The central shift in agentic IR is the evolving definition of ``information'' from \textit{static, pre-defined information items} to \textit{dynamic, context-dependent information states}. 
Information state refers to a particular information context that the user is right in within a dynamic environment, encompassing not only the acquired information items but also real-time user preferences, contextual factors, and decision-making processes. 
In such a way, traditional information retrieval, focused on acquiring relevant information items based on user queries, can be naturally extended to achieving the target information state given the user instruction, which thereby defines the agentic information retrieval. 
We systematically discuss agentic IR from various aspects, \ie, task formulation, architecture, evaluation, case studies, as well as challenges and future prospects. 
We believe that the concept of agentic IR introduced in this paper not only broadens the scope of information retrieval research but also lays the foundation for a more adaptive, interactive, and intelligent next-generation IR paradigm. 
\end{abstract}


\settopmatter{printacmref=false}


\maketitle

\section{Introduction}
\label{sec:intro}

Information retrieval (IR) is the process of obtaining relevant information from a large repository to meet user needs. 
It plays a crucial role in various real-world applications, such as web search, personalized recommendation, and online advertising. 
With the exponential growth of digital information in Internet applications, effective IR systems are essential for helping users navigate vast amounts of data efficiently and alleviating information overload~\citep{singhal2001modern,wang2017irgan,lin2023map}. 

\begin{figure*}[t]
    \centering
    \includegraphics[width=0.99\linewidth]{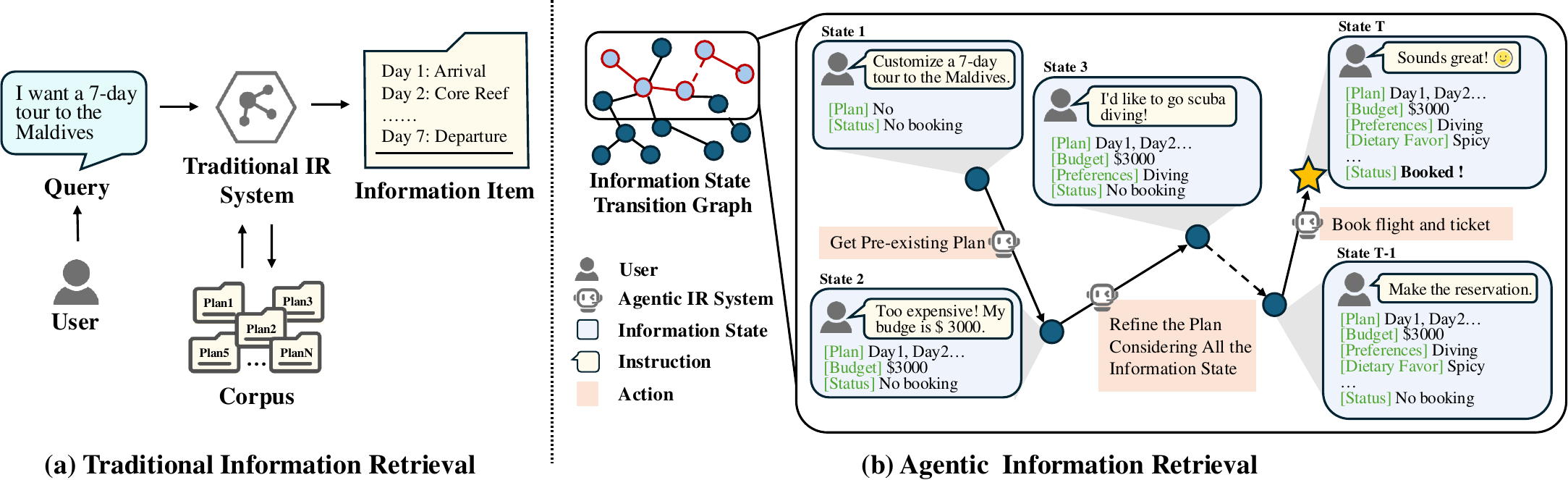}
    \caption{
    The next paradigm shifts from (a) traditional information retrieval to (b) agentic information retrieval. 
    We illustrate the core processes of both traditional IR and agentic IR with the example of online travel agency. 
    }
    \label{fig:illustration}
\end{figure*}

As shown in Figure~\ref{fig:illustration}(a), the traditional IR system follows a structured pipeline aimed at fulfilling user information needs.
The system first receives a query from the user, performs information filtering based on a pre-defined corpus, and finally returns the relevant information items to the user.
Such a process can be conducted for either one turn in a static mode (\eg, web search), or for multiple turns in an interactive mode (\eg, conversational recommendation). 
As a result, the user can iteratively refine the submitted query and eventually reach a satisfactory outcome. 

Despite the technical and business success, the traditional IR remains fundamentally constrained by its reliance on a static corpus. 
\textbf{The ``information to be retrieved'' is merely defined as the information item within a pre-defined corpus.}
For instance, the information items can be online websites for web search or products for e-commercial recommendation.  
Therefore, the existing IR systems merely filter and rank existing information items but lack the ability to manipulate, synthesize, or generate new contents. 
Taking the online travel agency as an example, traditional IR systems can only retrieve and rank existing travel plans from a pre-defined database, offering users a selective combination of available flights, hotels, and tour packages. 
However, they are unable to generate a fully customized itinerary that adapts to individual preferences and contextual factors such as budget constraints, weather conditions, or real-time ticket availability. 
Moreover, the existing IR systems cannot assist users in seamlessly placing orders and finalizing travel arrangements (\ie, task execution).

The emergence of large language models (LLMs) has fundamentally reshaped the nature of information retrieval \cite{zhu2023large,cai2024agentir,lin2023can,xi2024towards}. 
Furthermore, by wrapping LLMs as AI agents to interact between the user and the environment \cite{wang2024survey}, it is possible to revolutionize information retrieval with a broader information scope and deeper task integration. 
The systems empowered by AI agents are not only reactive for information item filtering, but also proactive for user intent reasoning, external tool usage, and ultimate task solving. 
With such a background, it is an opportune time to think about the next-generation IR architectures in the era of LLM-driven AI agents. 

To this end, in this perspective paper, we introduce the concept of \textbf{\textit{agentic information retrieval} (Agentic IR)}, a novel next-generation paradigm for information retrieval in the era of LLM-driven AI agents. 
The intrinsic foundation of agentic IR stems from the evolution of the ``information'' definition, which serves as the operating object for IR systems, compared with traditional IR as follows.
\begin{AIbox}{The New Definition of ``Information'' in IR}
In the era of LLM-driven AI agents, the definition of ``Information'' in IR is evolving from the \underline{information items} from the corpus to the \underline{information states} from the wild.
\end{AIbox}

Information state refers to a particular information context that the user is right in within a dynamic environment. 
With such a definition, the information item in traditional IR can be viewed as a special case of the information state. 
As shown in Figure~\ref{fig:illustration}(b), we still take the online travel agency as an example. 
The information state of the user not only includes the acquired pre-existing travel plans, but also involves the user's real-time preferences and iteratively refined arrangements, as well as the status of flight booking and hotel reservation. 
Each action taken by the agentic IR system, whether suggesting options, refining plans, or completing reservations, alters the information state by influencing the user’s cognitive knowledge and guiding the decision-making process. 

In such a way, information retrieval that regards acquiring relevant information items given the user query can be naturally extended to achieving the target information state given the user instruction. 
This shift from static information item acquisition to information state achieving in a dynamic environment defines our proposed agentic IR in the era of LLM-driven AI agents. 
As the interaction among the user, the IR system, and the external environment unfolds, the information state of the user is constantly updated and transferred from one to another, forming a \textit{information state transition graph} in the wild, which is illustrated in Figure~\ref{fig:illustration}(b).  
Based on the information state transition graph, we can compare the essential goal of agentic IR against traditional IR as follows:
\begin{AIbox}{Agentic IR vs. Traditional IR}
Agentic IR helps the user \underline{achieve the target information} \underline{state} on the state transition graph from the wild, while traditional IR merely helps the user \underline{acquire the preferred} \underline{information item} from a pre-defined corpus.
\end{AIbox}

In summary, we view agentic IR as a crucial and transformative evolution of traditional IR in the era of LLM-driven AI agents. 
It not only redefines the intrinsic nature of information (\ie, from information items to information states) and expands the operational boundaries of IR systems, but also provides a promising blueprint for the next-generation information entry. 
More than just fetching relevant data, IR systems should proactively help users access, integrate, refine, and act on information, seamlessly driving goal-oriented task executions.

The remaining part of this paper is organized as follows. 
In Section~\ref{sec:formualtion}, we present the task formulation of agentic IR. 
Then, we discuss the architecture design and evaluation protocol for agentic IR in Section~\ref{sec:architecture} and Section~\ref{sec:evaluation}, respectively. 
Next, in Section~\ref{sec:case study}, we give two case studies to further instantiate and explain the potential applications of agentic IR, \ie, life assistant and business assistant.
The key challenges and future prospects are provided in Section~\ref{sec:challenges}. 
Finally, we conclude this paper in Section~\ref{sec:con}.

\section{Task Formulation of Agentic IR}
\label{sec:formualtion}

In this section, we introduce the task formulation of agentic information retrieval. 
We first introduce the essential notations for agentic IR, and then formulate the optimization objective. 
Finally, we discuss the relationship of agentic IR with traditional IR.

\subsection{Essential Notations}

\subsubsection{Information State}

As discussed in Section~\ref{sec:intro}, since the capability boundary of IR systems has been fully extended by LLM-driven AI agents, the information object to be operated on is no longer simply organized as the static information item in a pre-defined corpus, but further involves the ever-evolving information context the user is right in within the dynamic environment. 

Formally, we can define the information state at a given time step $t$ during an agentic information retrieval (Agentic IR) process as  $s_t\in\mathcal{S}$.
Here, $\mathcal{S}$ is the global state space containing all the possible information states of the user during the interaction with the system. 
The state $s_t$ encapsulates the comprehensive cognitive state of the user towards the outside environment, \eg, the acquired information items, the LLM-generated responses, and the factual status like flight booking and online purchasing. 

\subsubsection{Action \& Policy}

The information state $s_t$ evolves as the agentic IR system takes actions to interact between the user and the environment. 
At each time step $t$, the agentic IR system would take an action $a_t\in\mathcal{A}$ according to the real-time feedback and instructions from the user. 
The global action space $\mathcal{A}$ is fairly diverse and closely related to the capabilities of back-end LLMs, including but not limited to acquiring relevant information items from various corpus, manipulating and integrating multi-source data, and making function calls for task execution with the external world.

We denote the overall policy of the agentic IR system as $\pi(a|s)$.
At each time step $t$, the policy takes as input the comprehensive user information state $s_t$ and generates the corresponding action $a_t$ to be delivered to the environment. 
Note that we include the per-step user instruction and real-time feedback in the information state, since the instruction given by the user can also reflect and affect his/her internal thoughts and preferences. 
The policy $\pi$ is typically implemented as an LLM-centered compounded system with several key components like the memory and planning modules, which will be further elaborated on in Section~\ref{sec:architecture}. 

\subsubsection{Environment \& State Trasition}

At each time step $t$, based on the policy $\pi(a|s)$, the agentic IR system generates the action $a_t$ and delivers it to the environment. 
The environment receives the action $a_t$ and gives the returns that can transfer the user information state $s_t$ to $s_{t+1}$. 
Formally, we formulate it as global state transition dynamics $p(s_{t+1}|s_t,a_t)$. 

\subsubsection{Information State transition graph}

Based on the notations above, as the interaction among the user, the system, and the environment unfolds, the user information state keeps transferring from one to another. 
By taking all the possible information states and actions into consideration, we can formulate the information state transition graph $\mathcal{G}=\{\mathcal{V},\mathcal{E}\}$, where $\mathcal{V}$ is the vertex set and $\mathcal{E}$ is the directed edge set. 
Each possible information state $s\in\mathcal{S}$ is regarded as a node on the graph, \ie, $\mathcal{V}=\mathcal{S}$. 
Each directed edge represents an action $a\in\mathcal{A}$ to make the potential transition between two states, \ie, $\mathcal{E}=\{(s_1,s_2)|s_2\sim p(\cdot|s_1,a),a\in\mathcal{A},\,s_1,s_2\in\mathcal{S}\}$.
The transition graph can cover all the possible contexts and states for the user, and is thereby established in the wild. 

\subsubsection{Reward Function} 
\label{sec:notation reward model}

Based on the information state transition graph introduced above, the agentic IR process for a user can be thereby denoted as an information state-action sequence $\tau=\{s_0,a_0,s_1,a_1,\dots,s_{T-1},a_{T-1},s_T\}$.
Here, $s_T$ serves as the terminated information state for the user interaction trajectory. 
The user can express his/her desired target information state $s_{*}$, \eg, the screening of a finalized travel plan with all the flights and tickets booked.
Then, we can build a verifier $r(s_{*}, \tau)$ as the reward function to estimate the performance of agentic IR system. 
The possible implementations for the verifier will be further discussed in Section~\ref{sec:evaluation}. 

\subsection{Optimization Objective}

By putting all the notations together, we can give the optimization objective of agentic information retrieval as follows.

The objective of an agentic IR system is to help the user navigate the state transition graph $\mathcal{G}$ efficiently, guiding the user from the initial state $s_0$ to a terminated state $s_T$ that best aligns with the desired target state $s_{*}$ expressed through the user instruction. 
In other words, the agentic IR system is required to help the user achieve the desired information state on the transition graph from the wild to satisfy his/her information needs. 

Formally, the core goal is to develop an optimal agent policy that maximizes the expectation of user satisfaction quantified by the reward function: 
\begin{equation}
\begin{aligned}
    \max_\pi~ & \mathbb{E}_{\tau}[r(s_*, \tau)]\\
    \text{s.t.~} & \tau=\{s_0,a_0,s_1,a_1,\dots,s_{T-1},a_{T-1},s_T\}, \\
    & a_t \sim \pi(\cdot|s_t),\; t=0,\dots, T-1, \\
    & s_{t+1}\sim p(\cdot|s_t, a_t),\; t=0,\dots, T-1, \\
\end{aligned}\label{eq:agent-ir-obj}
\end{equation}
where $\pi(a|s)$ is the policy to be optimized, $\tau$ is the agentic IR interactive trajectory, $s_{*}$ is the target information state expressed by the user, and $r(s_*, \tau)$ is the reward function.

\subsection{Relationship with Traditional IR}

Traditional IR can be viewed as a special case of agentic IR, when global spaces of information states and actions are restricted. 
Specifically, we can confine the information state to the set of retrieved information items presented to the user, rather than a dynamically evolving information context. 
Accordingly, the action space is also grounded to simply filtering from a pre-defined corpus without any further manipulation, \eg, calling a search engine API for web search, or invoking a recommendation model for e-commercial recommendation systems.
By constraining both the information state and action space, agentic IR reduces to traditional IR, making the latter a special case within the broader paradigm. 

Therefore, agentic IR represents a natural extension of traditional IR, driven by advancements in LLM-driven AI agents. ~\cite{wang2024towards}
While traditional IR has been instrumental in enabling efficient acquisition of relevant information items from a pre-defined corpus, it is fundamentally constrained by the filtering objective and static corpus of information items. 
In contrast, agentic IR transcends these limitations by integrating AI agents capable of multi-step reasoning, dynamic decision-making, and interactive task-solving with external tools to achieve user-desired target information states. 

This perspective highlights how expanding state representation and agent actions naturally evolves IR beyond simple acquisition for information items. 
We believe that the concept of agentic information retrieval proposed in this paper not only generalizes the research scope of information retrieval but also unlocks new possibilities, paving the way for a more adaptive, interactive, and intelligent next-generation IR paradigm. 


\section{Architecture of Agentic IR}
\label{sec:architecture}

In this section, we introduce the overall architecture of agentic IR systems, which are built based on LLM-driven AI agents. 
We first elaborate on the key components of one single AI agent and connect them to the field of information retrieval. 
Then we introduce the system-level design with AI agents, including both single-agent systems and multi-agent systems. 
Finally, we discuss key methods to develop and optimize the agentic IR system.

\subsection{Agent Components}

\begin{figure}[t]
    \centering
    \includegraphics[width=0.95\linewidth]{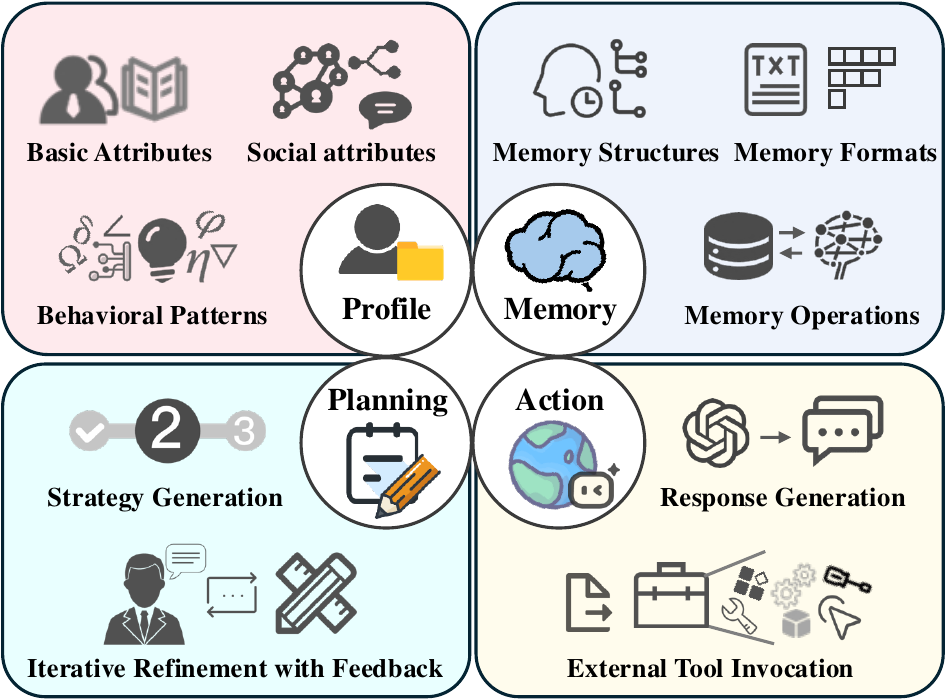}
    \caption{
    Four core components of the AI agent.
    }
    \label{fig:components}
\end{figure}

According to previous research~\citep{peng2025surveyllmpoweredagentsrecommender,singh2025agentic,wang2024survey,zhang2024generative}, each AI agent consists of four core components, \ie, profile module, memory module, planning module, and action module, which is shown in Figure~\ref{fig:components}.

\subsubsection{Profile Module}

The profile module defines the role-specific identity of the agent, influencing its behavior in information retrieval tasks. 
Unlike traditional IR systems that passively process queries, the agentic IR system operates as an autonomous agent that assumes specific roles, \eg, research assistant~\cite{openai2025deepresearch}, domain expert~\cite{ge2024openagi}, or investigative analyst~\cite{kapoor2024ai}. 
The profile module ensures that the agent could dynamically adapt to user needs and information contexts.

Agent profiles are typically embedded into the prompting structure of the system, guiding the agent’s reasoning style, response formulation, and interaction strategy~\cite{kapoor2024ai,hongmetagpt}. 
These profiles encompass multiple dimensions of characterization, \eg, basic attributes, behavioral patterns, and social information. 
The choice of profile characteristics depends on the applications. 
For instance, in stock-related information retrieval, an agent's decision-making biases and comprehensive thought processes might be explicitly modeled, whereas in an enterprise knowledge retrieval setting, the agent profile may emphasize formal tone, reliability, and citation accuracy.

By integrating these structured role definitions, the profile module ensures that the agent’s reasoning, task-solving, and response generation are aligned with its designated function, enabling consistent and context-aware interactions in complex agentic information retrieval scenarios. 

\subsubsection{Memory Module}
\label{sec:memory module}

The memory module serves as the cognitive backbone of an agentic IR system, enabling it to retain, recall, and reflect on past interactions to enhance future IR performance. 
The agentic IR system leverages memory to accumulate contextual knowledge, user-specific preferences, and environmental observations, allowing it to self-evolve and exhibit more consistent, rational, and effective behaviors.
Typically, the design of the memory module can be discussed in the following three key dimensions.
\begin{itemize}[leftmargin=10pt]
    \item \textbf{Memory Structures}. 
    Inspired by human memory processes, the memory module of an agent incorporates short-term memory and long-term memory~\cite{zhang2024survey}.
    The short-term memory can be implemented as the context window of a transformer-based LLM, and the long-term memory functions as an external knowledge base for persistent knowledge retention. 
    While short-term memory enables real-time adaptation during an interactive session for agentic IR, long-term memory allows the agent to maintain a historical knowledge base that can be efficiently queried and updated over extended interactions with users.
    \item \textbf{Memory Formats}. 
    The agent’s memory can be stored in various ways, including natural language memory, which retains explicit interaction histories in a human-readable format, and embedding memory, which encodes experiences as high-dimensional vector representations for fast reflection and semantic generalization~\cite{hatalis2023memory}. 
    The choice of format depends on the application. The natural language memory is well-suited for conversational agents requiring transparency, while the embedding memory is optimal for high-speed vector-based knowledge seeking. 
    \item \textbf{Memory Operations}. To effectively acquire, consolidate, and apply stored knowledge, the memory module supports three fundamental operations: (1) memory writing to persist the relevant information from past interactions; (2) memory reading to obtain relevant knowledge to adapt the agent's behaviors; (3) memory reflection to analyze, refine, and optimize the accumulated memories~\cite{hatalis2023memory}.
    Equipped with the memory module, the agent can learn from experience, refine its task-solving strategies over multiple interactions, and dynamically adjust their reasoning process to conduct more personalized, efficient, and contextually relevant information retrieval.
\end{itemize}

\subsubsection{Planning Module}

The planning module~\cite{huang2024understanding} is responsible for formulating and dynamically adapting multi-step strategies to guide the agent to achieve the target information state. 
The agentic IR system continuously plans, executes, and refines the interaction trajectories by reasoning over the current information state and environment constraints. 

The core characteristic of the planning module for agentic IR is that the planning has to incorporate the per-step real-time user feedback to iteratively adjust the internal policy. 
In complex information-seeking tasks, pre-defined plans often fail due to unforeseen constraints, evolving user needs, or dynamic information landscapes. 
For example, if an agent is assisting in an academic research task for reinforcement learning in robotics, an initial information retrieval strategy might prioritize survey papers for an overview of the research field. 
However, after reviewing the results, the user may express a further preference for recent applications in industrial automation (\eg, the user's information need evolves), requiring the agent to refine its planning to focus on more specific application papers.

By iteratively updating the planning for action strategies based on user feedback, the planning module enables adaptive and failure-resilient capabilities for agentic information retrieval. 
This human-like refinement process allows agentic IR systems to navigate the complex information state space effectively, ensuring that the final achieved state remains aligned with user expectations.

\subsubsection{Action Module}

The Action Module serves as the execution engine of the agentic IR system, translating the agent’s planned information-seeking strategies into concrete actions to be delivered to the environment. 
As the most downstream component, it directly interacts with the outside environment, which is influenced by the aforementioned profile, memory, and planning modules. 
Each time an action takes place, it would affect and trigger the user information state transition.

Apart from classical actions like chin-chatting and response generation for the user, the core action type of an LLM-driven AI agent is external tool usage (\ie, API calling)~\cite{qin2024toolllm,lin2024hammer}. 
The availability of tools determines the scope of actions the agent can execute. 
For example, only if the agent has access to the airline’s reservation API can it finalize the flight booking.
Without this reservation API, it may retrieve flight options but cannot complete the booking process for the user. 
Thus, the agent's functional capabilities are fundamentally constrained by its accessible external tools.
Moreover, the core functionality of traditional IR for relevant information item acquisition, such as the search engine for web search and recommendation model for e-commercial recommender systems, can also be wrapped as external APIs~\cite{xiong2024search, zhao2024recommender}. 
In such a way, agentic IR can be naturally viewed as a generalized extension of traditional IR, with core capabilities to not only acquire relevant information items but also proactively engage in reasoning and task execution.

\subsection{Agentic System Design}

After constructing the agent with key components introduced above, the design of agentic IR systems can be classified into two categories: single-agent system and multi-agent system.

\begin{figure}[t]
    \centering
    \includegraphics[width=0.95\linewidth]{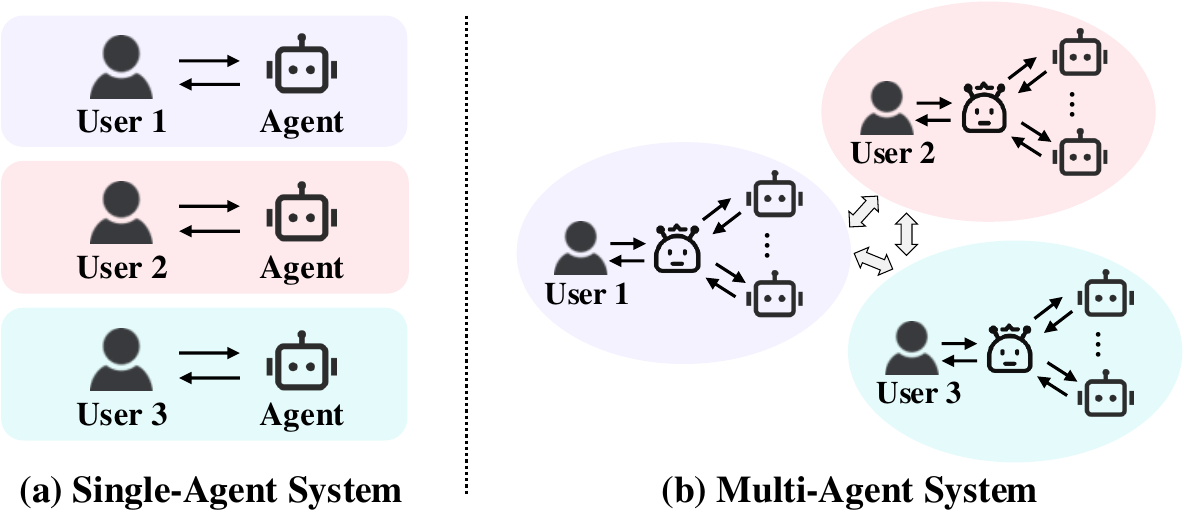}
    \caption{
    The design of single-agent and multi-agent systems for agentic information retrieval.
    }
    \label{fig:system design}
\end{figure}

\subsubsection{Single-Agent System}

The single-agent system relies on one single AI agent to fulfill the user's information needs. 
This setup is relatively simple and efficient since all the computational resources are dedicated to the operations of one agent. 
The agent follows a self-evolving pipeline to process instructions, interact with external tools, and iteratively refine the achieved information states for the user. 
For example, a legal research assistant could be designed as a single-agent system, where it fetches relevant case laws, summarizes key legal points, and refines results based on user feedback, ensuring that it remains efficient and targeted.

Despite its simplicity and efficiency, the single-agent design can struggle to handle more complex, multifaceted tasks, particularly when the problem requires different sources of personal information, diverse expertise or a high degree of specialization~\cite{li2024personal}. 
As the complexity of the user's request grows, the agent’s task-solving capabilities may become strained, leading to potential inefficiencies and suboptimal solutions. 
This constraint is one of the driving forces behind the design of multi-agent systems, where an orchestra agent distributes decomposed sub-tasks among specialized agents to handle more complex and diverse user needs.

\subsubsection{Multi-Agent System}

The multi-agent system extends the agentic IR framework by introducing multiple LLM-driven AI agents, which enhances the flexibility and scalability~\cite{yang2024multi}. 
As shown in Figure~\ref{fig:system design}(b), the multi-agent characteristic of the system can be discussed from the following two aspects:
\begin{itemize}[leftmargin=10pt]
    \item \textbf{Hierarchical multi-agent system for each user}. 
    Each user is served by multiple agents within his/her personalized system. 
    Typically, an orchestra agent acts as the primary interface with the user, decomposing complex tasks into sub-problems and distributing them to specialized downstream agents. 
    These downstream agents are delicately designed to handle specific sub-tasks, such as document retrieval and mathematical analysis. 
    For example, in a financial analysis system, an orchestra agent may analyze user queries, assign data extraction tasks to one agent, request market trend analysis from another, and compile findings into a comprehensive report for the user.
    \item \textbf{Collaborative multi-agent system among different users}. 
    The multi-agent characteristic also enables interactions among different individual users through their respective agent IR systems. 
    This aspect fosters enhanced information flow, allowing users to collaboratively refine their states, share insights, and build collective knowledge repositories. 
    For instance, in academic research, multiple scholars may use their own agentic IR systems to coordinate literature reviews, cross-reference findings, and collectively refine their research direction.
\end{itemize}
By incorporating multiple agents, agentic IR systems gain robustness, scalability, and adaptability, making it well-suited for diverse and complex agentic IR scenarios. 

\subsection{Key Methods for Optimization}

Based on the agent components and agentic system design above, here we briefly introduce the key methods to improve the performance of agentic IR, \ie, the objective in Eq.~\ref{eq:agent-ir-obj}. 
The methods below are all designed for LLMs, which serve as the centers of agentic IR systems.
\begin{itemize}[leftmargin=10pt]
    \item \textbf{Prompt Engineering}. 
    Prompts are the task-based language token input to the LLM to enable its ability for the task~\citep{liu2023pre}. 
    For an LLM-driven AI agent, the prompt is a human-controllable way to set its hidden state in comparison to the model parameters, including but not limited to the chain-of-thought prompting~\citep{wei2022chain} and reflection techniques~\citep{shinn2024reflexion}.
    \item \textbf{Retrieval-Augmented Generation (RAG)}. 
    The task-related demonstrations play a crucial role in LLM-based applications \cite{izacard2021leveragingpassageretrievalgenerative,du2024codegrag}. 
    In agentic IR, the demonstrations can be retrieved at either the action level or the trajectory level~\citep{zheng2024synapse,zhou2024trad}. 
    
    \item \textbf{Supervised Fine-tuning (SFT)}. As a basic method for fine-tuning LLMs, SFT can be seamlessly adapted to agentic IR tasks, where the successful historical trajectories are used as the training data with each step of generated action as the label to fit. 
    SFT corresponds to the behavioral cloning imitation learning method in reinforcement learning. 
    Despite simplicity, SFT does not directly optimize the objective formulated in Eq.~\ref{eq:agent-ir-obj}.
    \item \textbf{Pairwise Preference Learning}. As one step further based on SFT, fine-tuning LLMs based on a preference objective over a pair of outputs can improve the performance of the agentic IR system~\citep{rafailov2024direct}. 
    Note that such methods are to some extent similar to the pairwise learning to rank techniques in traditional IR~\citep{burges2005learning}, \ie, given the current information state, which action can make the next state more desirable to the user.
    \item \textbf{Reinforcement Fine-tuning (RFT)}. Regarding the interaction with environment as a Markov decision process, reinforcement learning methods (\eg, PPO~\citep{schulman2017proximal}) directly optimize the objective in Eq.~\ref{eq:agent-ir-obj}, given the reward signal from the environment or human feedback (RLHF)~\citep{ouyang2022training}. 
    Compared with SFT and preference learning, RFT usually requests larger computational resources to explore the environment, accumulate experience data, and optimize the model performance~\citep{luong2024reft,christianos2023pangu}.
    \item \textbf{Reward Modeling}. 
    As a judge of the terminated or intermediate information states in the process, the reward function modeling is crucial to enable RFT or search-based decoding techniques in complex agentic IR tasks. 
    Referring to recent advances in math reasoning problems~\citep{uesato2022solving,luo2024improve}, process reward models are essential to provide fine-grained per-step supervisions through the whole trajectory, thereby promoting the performance of the agent.
    \item \textbf{Complex Reasoning}. 
    For non-trivial IR tasks, the agent needs to perform task planning and complex reasoning before taking actions. 
    The recent success of OpenAI o1~\citep{openai2024o1} and DeepSeek-R1~\citep{guo2025deepseek} indicates the great potential of a strong reasoner with explicit chained thoughts for improving the agent's task-solving performance to satisfy user information needs.
    \item \textbf{Multi-Agent Systems (MAS).} 
    A multi-agent system contains multiple homogeneous or heterogeneous agents, each of which could be equipped with a special role or resources. 
    With proper mechanisms, the team of agents manages to coordinate to achieve remarkable collective intelligence~\citep{chen2023agentverse,li2024more} for agentic IR tasks.
\end{itemize}

\section{Evaluation Protocol of Agentic IR}
\label{sec:evaluation}

Traditional IR typically focuses on ranking tasks to return a list of information items for the user, and therefore evaluates system performance \textbf{at the list level} by considering factors like relevance, utility, diversity, and fairness. 
However, with the shift to agentic IR, the evaluation must now occur \textbf{at the trajectory level}. 
Rather than just acquiring a ranked list of information items, agentic IR systems help users navigate towards a desired information state through a series of interactive actions and states. 
This requires a new evaluation protocol between the desired target state and the entire interaction trajectory. 

As discussed in Section~\ref{sec:notation reward model}, to evaluate the system's performance in such a context, we introduce a verifier as the reward function. 
This verifier, denoted as $r(s_{*},\tau)$, compares the desired target information state $s_{*}$ with the entire trajectory $\tau=\{s_0,a_0,\dots,a_{T-1},s_T\}$. 
This trajectory-based evaluation allows us to holistically assess both the process and the outcome of the agent’s actions. 
Specifically, there are three primary metric types with such a formalized verifier: utility-oriented, efficiency-oriented, and ethics-oriented metrics. 

\subsection{Utility-Oriented Metrics}

Utility-oriented metrics primarily focus on the effectiveness of the agentic IR system in fulfilling the user's information needs. 
In agentic IR, the task success rate (TSR) is one of the most critical metrics in this category. 
The TSR measures the similarity between the achieved final information state $s_T$ and the desired target information state $s_{*}$ in either a binary manner (exactly matched or not) or a numerical manner (win partial scores if not exactly matched). 
This metric directly reflects the user satisfaction and indicates the degree to which the system has successfully navigated the user towards their intended goal on the information state transition graph. 
The measurement of similarity can be conducted by incorporating rule-based judgment, human annotation, or AI feedback based on powerful foundation models like GPT-4o. 

Note that there is an alternative choice to compare the trajectory taken by the agent to the ideal ground-truth trajectory recorded in the offline dataset. 
However, this comparison has inherent limitations and biases, since there must exist multiple distinct trajectories on the transition graph that can lead to the same final information states. 
Therefore, while trajectory comparison can provide insight into the system's capabilities through the fine-grained process evaluation, it should be treated with caution.

\subsection{Efficiency-Oriented Metrics}

While agentic IR systems offer more generalized solutions and better task-solving capabilities, they often come at a higher computational cost compared to traditional IR. 
Hence, the efficiency-oriented metrics are designed to evaluate how effectively the system can balance its sophisticated capabilities with the computational resources it consumes. 
We discuss three key metrics in this category:
\begin{itemize}[leftmargin=10pt]
    \item \textbf{Run Time}. The inference time taken by the agentic IR system to complete the entire information retrieval process, including all interactions between the user and the system. 
    We can also record the per-step run time of the system at each turn.
    \item \textbf{Token Consumption}. The total number of tokens processed by the LLM-centered agentic IR system during the interaction. 
    Larger token consumption indicates potentially longer run time and higher resource cost. 
    Similarly, per-step token consumption at each turn can be recorded.
    \item \textbf{Number of Interaction Turns}. 
    A lower number of interactions (\ie, turns) between the user and the system denotes the better efficiency of the system. 
    Moreover, the number of interaction turns is also a valuable indicator of the personalization capabilities. 
    A system that can guide a user to the desired information state with fewer interactions is likely to have a better understanding of the user preferences, demonstrating a higher degree of personalization and user modeling.
\end{itemize}

\subsection{Ethics-Oriented Metrics}

As agentic IR systems become more integrated into users' real-life decision-making and task-executing processes, it is crucial to assess their ethical implications. ~\cite{dai2024bias}
Ethics-oriented metrics evaluate the safety, fairness, and social good characteristics of agentic IR systems. 
These metrics focus on factors such as:
\begin{itemize}[leftmargin=10pt]
    \item \textbf{User-Friendliness}. 
    Whether the interaction trajectory is intuitive, respectful, and transparent. 
    We have to ensure that the agentic IR system does not lead the user to misleading or harmful paths during the information seeking and manipulation. 
    \item \textbf{Safety and Security}. 
    The extent to which the system ensures user data privacy and operates within secure boundaries. 
    For example, whether the agentic IR system is robust to prompt or even code injection attacks during the external API callings. 
    \item \textbf{Bias and Fairness}. Whether the system exhibits biases during its interaction with the user, which might potentially lead to unfair treatment of certain users or groups. 
\end{itemize}

\section{Case Studies}
\label{sec:case study}

\begin{figure*}
    \centering
    \includegraphics[width=1.0\linewidth]{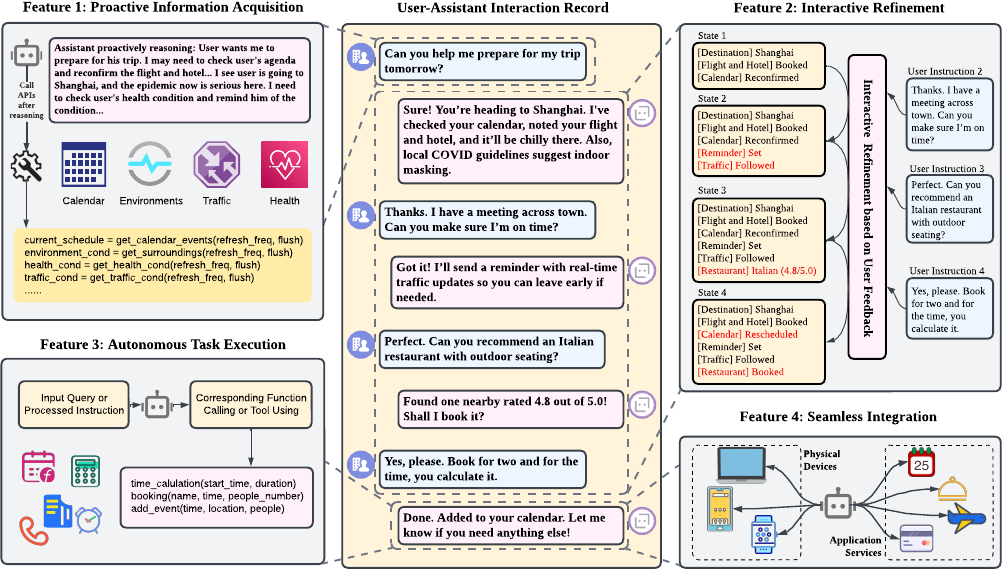}
    \caption{
    The case study of agentic IR in life assistant scenarios, which conveys four key features.}
    \label{fig:lifeassistant}
\end{figure*}

In this section, we give two case studies of agent IR, \ie, life assistant and business assistant. 
Due to the extended capabilities of external tool usage and task execution, the agent IR would play more like an assistant to help users access and manipulate the information with a certain level of autonomy. 
In the case study of life assistant, we would like to illustrate the key features of agentic IR, while in the case study of business assistant, we further demonstrate the core stages for the user-agent interaction.

\subsection{Life Assistant}

In recent years, with the emerging techniques of agentic IR, life assistants have evolved from simple voice-activated chatbots into sophisticated systems capable of supporting users across a wide range of daily tasks. 
Agentic IR can not only gather and deliver information but also  proactively support planning, decision-making, and task execution with a deep understanding of the user’s needs, contexts, and preferences. 
This shift enables life assistants to act as active, autonomous agents that adapt seamlessly to a user's real-world life, offering guidance and assistance.

Agentic IR capabilities are already presented in real-world products like Apple’s ecosystem, where Apple Intelligence\footnote{\href{https://www.apple.com/apple-intelligence/}{https://www.apple.com/apple-intelligence/}} powers advanced assistant features across devices such as iPhone, iPad, and Mac~\citep{apple-intelligence-foundation-language-models}. Apple Intelligence enhances user experience by seamlessly integrating with apps, services, and smart devices, embodying the proactive and contextual characteristics of agentic IR. Other life assistants, such as Google Assistant\footnote{\href{https://assistant.google.com/}{https://assistant.google.com/}}, Oppo Breeno, and Huawei Celia\footnote{\href{https://consumer.huawei.com/en/emui/celia/}{https://consumer.huawei.com/en/emui/celia/}}, operate across diverse platforms, including smartphones, smart home devices, and wearables. 
These assistants empower users with convenient control over both digital and physical environments, enabling them to make informed plans and adjustments anytime, anywhere~\citep{li2024personalllmagentsinsights}.

As shown in Figure~\ref{fig:lifeassistant}, we give the case study by considering the following scenario: Jane is a busy professional who uses a life assistant integrated into her smartphone and other devices. Agentic IR allows her assistant to anticipate her needs, gather relevant information, and autonomously perform tasks without constant user intervention. 
Through such a case study, we would like to discuss the following four key features conveyed by the agentic-IR-empowered life assistant.

\textbf{Proactive Information Acquisition}. 
As shown in Figure~\ref{fig:lifeassistant}, the initially received instruction from Jane for the trip planning is actually vague and missing various important information facts like the destination. 
Hence, the assistant would first conduct in-depth reasoning to infer Jane's personalized preferences and potential needs, and then proactively invoke external tools to acquire the possibly related information in advance. 
As a result, the assistant goes through Jane's calendar and obtains the key information that Jane is heading for Shanghai. 
This presents the key feature of the agentic IR system to conduct complex reasoning about the user's implicit preference and potential needs, thereby proactively gathering the relevant information in advance.



\textbf{Interactive Refinement}.
Agentic IR enables the assistant to adapt to the user's real-time feedback by iteratively performing different actions based on both explicit per-step instructions and passive contextual cues, allowing continuous updates and refinements of the user information state. 
As shown in Figure~\ref{fig:lifeassistant}, after Jane requests a restaurant recommendation, the assistant iteratively clarifies her preferences (\eg, cuisine and view) via multiple rounds of interactions to better align with her preferences and needs. 
By adapting to both explicit responses and situational context, the assistant effectively helps Jane progress toward the desired target information state, underscoring agentic IR’s capacity for flexible, accurate assistance with interactive refinements.

\textbf{Autonomous Task Execution}.
Beyond simply gathering information in traditional IR, agentic IR enables the assistant to autonomously execute tasks, such as booking a dinner reservation or setting reminders. 
When Jane’s life assistant books a restaurant and returns the reservation feedback, she reaches an information state where the booking is completed and confirmed in her calendar. 
This autonomous task-executing capability alleviates Jane from cognitive burden, enabling her to concentrate on higher-priority tasks while the assistant can help her effortlessly transist to the desired target information state.

\textbf{Seamless Integration}. 
Another important feature is that the life assistant empowered by agentic IR can be seamlessly integrated into various physical devices and online services. 
As shown in Figure~\ref{fig:lifeassistant}, although the interactive conversation happens on Jane's mobile phone, the back-end life assistant can seamlessly update and synchronize all the information across different physical devices. 
Moreover, since the agent interacts with the environment based on natural language and standard tool interface, the assistant is able to set up the calendar reminder for different devices like the smart watch via the external API calls, ensuring that her arrival time and thermostat settings are in harmony. 
Each implicit integration step across different devices and services can change the user information state of Jane, keeping the physical environment status (\eg, restuarant reservation and calendar arrangement) aligned with her personal schedule and individual preference.

In summary, agentic IR represents a fundamental shift in how life assistants interact with users and how they can change the lifestyle of human beings for information accessing and manipulation. 
By anticipating needs, understanding context, and performing tasks autonomously, agentic IR makes life assistants not just more useful but indispensable in facilitating our daily lives. 
The proactive nature of agentic IR systems--along with their ability to integrate multiple sources of information, learn from interactions, and act independently--enables them to provide a uniquely tailored and efficient user experience for the next-generation IR paradigms.

\subsection{Business Assistant}

Business assistants are designed to support enterprise users by providing relevant business knowledge and insights based on various documents and data sources. 
Equipped with LLM-driven AI agents, business assistants go beyond passively acquiring relevance information pieces, proactively engaging in intention recognition, information manipulation, and response generation. 
In this way, business assistants can address a broad spectrum of business-related requirements, from financial analysis to marketing strategies, empowering users to make wise decisions.

\begin{figure*}
    \centering
    \vspace{10pt}\includegraphics[width=1\linewidth]{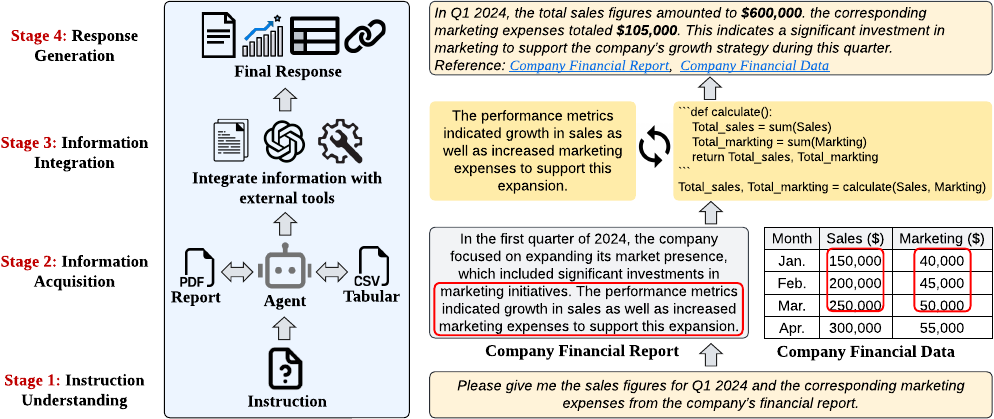}
    \caption{The case study of agentic IR in business assistant scenarios, which consists of four key stages.}
    \label{fig:documentqa}
\end{figure*}

Currently, there are already several business assistants powered by agentic IR in real-world applications, such as Microsoft 365 Copilot\footnote{\href{https://www.microsoft.com/en-us/microsoft-365/copilot/copilot-for-work}{https://www.microsoft.com/en-us/microsoft-365/copilot/copilot-for-work}}, Notion AI \footnote{\href{https://www.notion.so/product/ai}{https://www.notion.so/product/ai}}, and IBM watsonx\footnote{\href{https://www.ibm.com/watsonx}{https://www.ibm.com/watsonx}}. 
Typically, as illustrated in Figure~\ref{fig:documentqa}, the workflow of a business assistant consists of four key stages: instruction understanding, information acquisition, information integration, and response generation. 
Each of these stages is crucial for delivering tailored business insights via in-depth reasoning and task execution by the AI agent.

\textbf{Instruction Understanding}.
Given a business-related instruction, the agent, as the core of the business assistant, first attempts to discern the user's intent. 
For complex instructions, the agent can generate thoughts to break the instruction into smaller, manageable steps, facilitating multi-step reasoning and planning. 
In addition, with the memory module introduced in Section~\ref{sec:memory module}, the agent can use previous interaction records as contextual memory, deepening its capabilities to better understand and anticipate the user's preferences and needs.

\textbf{Information Acquisition}.
After identifying the user intents, the agent starts to acquire relevant data from both internal and external knowledge bases, ensuring that it extracts the most pertinent information. 
Given the variety of document formats (\eg, PDFs, tables, and images), the agent can utilize tools such as OCR for scanned text or SQL for structured data. 
Moreover, semantic search could enable the agent to go beyond keyword matching, ensuring that acquired information aligns more closely with the user's potential intent.

\textbf{Information Integration}.
In many cases, the acquired relevant information is scattered across multiple sections or even different documents. 
To construct a comprehensive response, the agent must combine and fuse the multi-source information. 
By generating internal thoughts, the agent can establish logical connections between disparate pieces of information and perform complex reasoning, gradually working toward the final information state. 
Additionally, specialized tools can also assist in this process by enabling capabilities such as executing mathematical calculations and filtering out redundant information. 

\textbf{Response Generation}.
Finally, the agent generates a response and presents it to the user, thereby transferring the user information state. 
The response may be presented in multiple formats, including plain text, tables, visualized charts, etc. 
Moreover, for transparency and trustworthiness, responses can also be linked back to their original source documents, allowing users to trace how the information was derived. 
Given the response from the business assistant, the user can either terminate this round of agentic information retrieval process, or give further instructions to refine the outcome, heading for a more desired target information state. 

The application of business assistant is continuously evolving with advancements in agentic IR and growing market demand. 
Key trends include enhanced contextual understanding and multi-step reasoning, enabling business assistants to comprehend and execute more complex instructions. 
Furthermore, since the business data keeps changing and updating all the time, business assistants have to retrieve and integrate real-time information from ever-evolving data sources. 

\section{Challenges and Future Prospects}
\label{sec:challenges}

Agentic information retrieval (Agentic IR) introduces a novel paradigm with the potential to revolutionize how we seek and interact with information. 
However, agentic IR is still in its nascent stages, and several critical challenges must be overcome for it to reach its full potential. 
In this section, we discuss the key challenges and future prospects for agentic IR as follows:
\begin{itemize}[leftmargin=10pt]
    \item \textbf{Exploration-Exploitation Tradeoff}. 
    The effectiveness of agentic IR depends heavily on high-quality interaction data, which is shaped by a combination of user instructions, the agent policy, and the dynamics of the environment. 
    Gathering such data is complex due to the tradeoff between exploration and exploitation during the agentic IR system's decision-making. 
    While exploration allows the agent to discover new possibilities, exploitation focuses on optimizing known behaviors. 
    Balancing these two strategies is critical for ensuring broad and diverse data coverage. 
    \item \textbf{Model Training}. 
    In traditional IR systems, model training typically involves optimizing a few specific models, such as retrieval models and ranking models, each focused on a particular aspect of the IR pipeline. 
    However, agentic IR represents a much more complex and composite system that requires training and fine-tuning across interdependent modules. 
    This complexity casts significant challenges over model training, which has to not only focus on optimizing individual components but also consider their coordination to achieve coherent performance. 
    \item \textbf{Inference Efficiency.}
    Due to the large parameter size and autoregressive nature of LLMs, LLM-driven agentic IR is computationally expensive and time-consuming to run in real-world applications. 
    Balancing inference speed, resource consumption, and model performance remains a key challenge for large-scale agentic IR systems. 
    As for potential solutions, techniques like model pruning, knowledge distillation, and hybrid inference strategies like speculative decoding could help reduce the computational overhead while maintaining the overall accuracy. 
    \item \textbf{Safety and Alignment}. 
    Safety is a critical concern for agentic IR systems, as these agents directly interact with the outside environment and influence the information states to which users are exposed. 
    Unlike traditional IR that merely acquires information items from a pre-defined corpus, agentic IR is allowed to take actions and execute tasks by invoking external APIs, which can affect the environment and the user’s journey. 
    As such, ensuring that these agents behave safely and align with human values is a fundamental challenge. 
    Recent work on the ``world model + verifier'' framework~\cite{ji2023ai,dalrymple2024towards} provides a promising solution, where the agent simulates potential outcomes and verifies the consequences of its actions within a simulative world model.
    \item \textbf{Human-Agent Interaction}.
    Proactively interacting with users is one of the most distinctive aspects of agentic IR.
    The agent must continuously update its understanding of user intent and preferences, while also adapting to ambiguities, contradictions, or incomplete inputs from users. 
    Note that the interaction patterns between users and agentic IR systems are no longer limited to texts on screens, but can be further extended to multi-modal manners like voice dialogue or even video chatting according to different product forms. 
\end{itemize}

\section{Conclusion}
\label{sec:con}

This perspective paper introduces the concept of agentic information retrieval (\textbf{Agentic IR}), a next-generation paradigm shift that redefines IR in the era of LLM-driven AI agents. 
Unlike traditional IR, which reactively acquires relevant information items from a pre-defined, agentic IR operates on the user information states, enabling AI agents to reason over user intent, interact with environments, and proactively execute tasks. 
We provide systematic discussion about agentic IR from various aspects, \ie, task formulation, architecture, evaluation protocol, case studies, as well as challenges and future prospects. 
We believe that the concept of agentic IR proposed in this paper not only generalizes the research scope of information retrieval but also paves the way for a more adaptive, interactive, and intelligent next-generation IR paradigm. 


\section*{Acknowledgments}
We thank the insightful discussions with Yifan Zhou, Jie Fu, Jun Wang, Ying Wen, Zheng Tian, Qiuying Peng, and Grace Hui Yang.

\bibliographystyle{ACM-Reference-Format}
\bibliography{sample-base}

\end{document}